\documentclass[12pt]{article}
\usepackage{epsfig}
 \newcommand{\app}[1]{Appendix~\ref{sect.#1}}
 \newcommand{\sect}[1]{\S\ref{sect.#1}}      
 \newcommand{\eq}[1]{Eq.~(\ref{eq.#1})}	
 \newcommand{\bareeq}[1]{(\ref{eq.#1})}	
 \newcommand{\fig}[1]{Fig.~\ref{fig.#1}}

 \newcommand{\sectlabel}[1]{\label{sect.#1}}
 \newcommand{\eqlabel}[1]{\label{eq.#1}}

\newcounter{eqletter} 

\newenvironment{mathletters}{
\setcounter{eqletter}{0}
\refstepcounter{equation}

}{}

\newenvironment{equationGroup}{
\addtocounter{equation}{-1} 
\stepcounter{eqletter}
\begin{equation}
}{\end{equation}}

\newcommand{\figdef}[3]{
\begin{figure}[!htb]
 \centering\leavevmode#2%
 \caption{\small #3}
 \label{fig.#1}
\end{figure}                 }

\newcommand{\ones}[1]{{| #1 |}}

\newcommand{\bitA}{\wedge}

\newcommand{\W}{\hat{W}}

\newcommand{\Nb}{ { N_{\rm better} } }

\newcommand{\NStart}{ { N_{\rm start} } }
\newcommand{\cStart}{ { c_{\rm start} } }

\newcommand{\floor}[1]{ { \left \lfloor #1 \right \rfloor } }

\begin{document}
\title{Local Search Methods for Quantum Computers}

\author{
Tad Hogg \\
Xerox Palo Alto Research Center \\
3333 Coyote Hill Road, Palo Alto, CA 94304 \\
hogg@parc.xerox.com
	\and
Mehmet Yanik \\
Dept. of Electrical Engineering \\
MIT \\
Cambridge, MA 02139 \\
fatih@mit.edu
}

\maketitle

\begin{abstract}
  Local search algorithms use the neighborhood relations among
  search states and often perform well for a variety of NP-hard
  combinatorial search problems.  This paper shows how quantum
  computers can also use these neighborhood relations. An example of
  such a local quantum search is evaluated empirically for the
  satisfiability (SAT) problem and shown to be particularly
  effective for highly constrained instances.  For problems with an
  intermediate number of constraints, it is somewhat less effective
  at exploiting problem structure than incremental quantum methods,
  in spite of the much smaller search space used by the local
  method.
\end{abstract}

\section{Introduction}

Combinatorial search problems are among the most difficult
computational tasks: the time required to solve them often grows
exponentially with the size of the problem~\cite{garey79}.  Many such
problems have a great deal of structure, allowing heuristic methods to
greatly reduce the rate of exponential growth.  Quantum
computers~\cite{benioff82,bernstein92,deutsch85,deutsch89,divincenzo95,feynman86,lloyd93}
offer a new possibility for utilizing this structure with {\em quantum
parallelism}, i.e., the ability to operate simultaneously on many
classical search states, and {\em interference} among different paths
through the search space.

While several algorithms have been
developed~\cite{shor94,boyer96,grover96,hogg95d,hogg97}, the extent to
which quantum searches can improve on heuristically guided classical
methods for NP searches remains an open question.  Even if quantum
computers are not useful for {\em all} combinatorial search
problems, they may still be useful for many instances encountered in
practice. This is an important distinction since typical instances of
search problems are often much easier to solve than is suggested by
worst case analyses, though even typical costs often grow
exponentially on classical machines.  The study of the average or
typical behavior of search heuristics relies primarily on empirical
evaluation. This is because the complicated conditional dependencies
in search choices made by the heuristic often preclude a simple
theoretical analysis, although phenomenological theories can give an
approximate description of some generic
behaviors~\cite{hogg94b,kirkpatrick94,monasson96}.

The simplest algorithms are the unstructured methods, such as
generate-and-test, which amount to a random search among the states or
a systematic enumeration of them without any use of prior results to
guide future choices. An analogous unstructured quantum search can
improve on this classical method~\cite{grover96}.

One structured approach to search {\em builds} solutions incrementally
from smaller parts. These methods exploit the fact that in many
problems the small parts can be tested for consistency before they are
expanded to larger ones. When a small part is found to be
inconsistent, all possible extensions of it will also be inconsistent,
allowing an early pruning of the search. In such cases, the search
backtracks to a prior decision point to try a different incremental
construction.  Analogies with this approach form the basis of
previously proposed structured quantum searches~\cite{hogg95d,hogg97}.

A second approach takes advantage of the clustering of solutions found
in many search problems. That is, instead of being randomly
distributed throughout the search space, the states have a simple
neighborhood relationship such that states with a few or no conflicts
tend to be near other such states. This neighborhood relationship is
used by {\em repair} or {\em local} searches. Starting from a random
state, they repeatedly select from among the current state's neighbors
one that reduces the number of conflicts with the constraints. Such
searches can become stuck in local minima but are often very
effective~\cite{minton92,selman92}. The problem of local minima can be
addressed by allowing occasional changes that increase the number of
conflicts~\cite{kirkpatrick83}. By comparison with incremental
methods, local searches operate in a smaller search space, i.e., they
have no need to consider the many small parts that might be composed
into a full solution. On the other hand, they also disregard any
information from these small parts that can guide the search toward a
solution.

In this paper, we present a new quantum search algorithm based on the
neighborhood structure of the satisfiability problem.  Such structure
forms the basis of many local classical methods but has not yet been
applied to quantum computation. Specifically the following two
sections describe the satisfiability problem used to illustrate our
algorithm and the ingredients of quantum programs.  We then describe a
quantum algorithm analogous to classical local search methods and
evaluate its behavior.  Finally some open issues are discussed,
including a variety of ways this approach can be extended.  The
appendices derive the properties of the algorithm's quantum operation
and discuss approaches to the classical simulation of the algorithm.

\section{The Satisfiability Problem}\sectlabel{sat}

NP search problems have exponentially many possible states and a
procedure that quickly checks whether a given state is a
solution~\cite{garey79}.  Constraint satisfaction problems
(CSPs)~\cite{mackworth92} are an important example.  A CSP consists of
$n$ variables, $V_1,\ldots,V_n$, and the requirement to assign a
value to each variable to satisfy given constraints. Searches examine
various {\em assignments}, which give values to some of the
variables. {\em Complete} assignments have a value for every variable.

One important CSP is the satisfiability problem (SAT), which consists
of a propositional formula in $n$ variables and the requirement to
find a value (true or false) for each variable that makes the formula
true. This problem has $N=2^n$ complete assignments. For
$k$-SAT, the formula consists of a conjunction of clauses and each
clause is a disjunction of $k$ variables, any of which may be
negated. For $k \geq 3$ these problems are NP-complete. An example of
such a clause for $k=3$, with the third variable negated, is $V_1$ OR
$V_2$ OR $\overline{V_3}$, which is false for exactly one assignment
for these variables: $\{V_1={\rm false}, V_2={\rm false}, V_3={\rm
true}\}$.  With the formula expressed as a conjunction of clauses, a
solution must satisfy every clause. Thus we can view each clause as a
constraint that adds one conflict to all complete assignments that
include assignments to the variables in the clause that make it false.
By selecting the clauses to be distinct, the number of clauses $m$
corresponds to the number of constraints in the problem.

The assignments for SAT can also be viewed as bit strings with the
correspondence that the $i^{th}$ bit is 0 or 1 according to whether
$V_i$ is assigned the value false or true, respectively. In turn,
these bit strings are the binary representation of integers, ranging
from 0 to $2^n-1$.  For definiteness, we associate the value of
variable $V_i$ with the bit in the integer corresponding to $2^{i-1}$,
i.e., $V_1$ corresponds to the least significant bit. Each clause in
the a $k$-SAT formula specifies a combination of $k$ bits that cause a
conflict in any assignment containing all of them. For example, the
clause $V_1$ OR $V_2$ OR $\overline{V_3}$ introduces a conflict in all
assignments for which bits 1 and 2 are 0 and bit 3 is 1, i.e.,
assignments whose integer representation ends with the bit string 100.

For bit strings $r$ and $s$, we define $\ones{s}$ to be the number of
1-bits in $s$ and $r \bitA s$ to be the bitwise AND operation on $r$
and $s$. Thus $\ones{r \bitA s}$ counts the number of 1-bits both
assignments have in common, and can also be viewed as a bitwise dot
product of the two assignments considered as vectors of bits. We also
use $d(r,s)$ as the Hamming distance between $r$ and $s$, i.e., the
number of positions at which they have different values. These
quantities are related by
\begin{equation}\eqlabel{hamming}
d(r,s) = \ones{r} + \ones{s} - 2 \ones{r \bitA s}
\end{equation}

An example with $n=2$ and two constraints,
each involving a single variable, is $V_1 \neq 1$ and $V_2 \neq 1$.
This problem has a unique solution: $V_1=0,
V_2=0$, an assignment with the bit representation 00.
This problem is also an instance of 1-SAT with the propositional formula
$\overline{V_1}$ AND $\overline{V_2}$.
The $N=2^n=4$ complete assignments have the bit representations
00, 01, 10, and 11.

The search algorithm presented below uses the average number of
conflicts in an assignment, $c_{\rm avg}$.  Each of the $m$ distinct
clauses in a $k$-SAT problem introduces a single unique conflict in
all complete assignments whose values for the variables specified in
the clause match those of the clause.  Since there are $2^{n-k}$ such
assignments, the total number of conflicts in all assignments is $m
2^{n-k}$. Thus the average number of conflicts in an assignment is
\begin{equation}\eqlabel{conflicts}
c_{\rm avg} = m / 2^k
\end{equation}

In evaluating average behavior of search algorithms, a common class of
problems used is the ensemble of random $k$-SAT problems. This
ensemble is specified by values for the number of variables $n$, the
size of the clauses $k$ and the number of distinct\footnote{This
ensemble differs slightly from other studies where the clauses are not
required to be distinct.}  clauses $m$.
 
In using random $k$-SAT to test our algorithm, we restrict attention
to those with at least one solution. Specifically, to generate ``random
soluble problems'', a set of $m$ distinct clauses is randomly selected
from the ${n \choose k} 2^k$ possible clauses~\cite{nijenhuis78}. The
resulting SAT problem is then solved with a classical backtrack search
to see if it has any solutions. Only those instances with solutions are
retained.

Soluble problems are very rare among randomly generated instances with
many constraints. Thus to allow comparison over the full range of
$m/n$ we generate a slightly different ensemble, random problems with
a prespecified solution. To generate these problems, a random
assignment is first selected to be a solution. Then, $m$ distinct
clauses are selected from among those that do not conflict with that
prespecified solution.  For $k$-SAT, this leaves
\begin{equation}\eqlabel{mMax}
m_{\rm max} = {n \choose k} (2^k - 1)
\end{equation}
possible clauses to select among,
because the prespecified solution precludes any clause that exactly
matches the assignments in that solution.

Compared to random soluble problems, using a prespecified solution
increases the likelihood of problems with a larger number of
solutions, resulting in slightly easier problems both classically and
for the quantum search algorithm described in this paper. However, at
the extremes of $m=0$ and $m=m_{\rm max}$, both methods give the same
results. Furthermore, when $m/n$ is fairly large, almost all soluble
problems have only a single solution, resulting in very little
difference between these two methods of generating soluble problems.

\section{Quantum Computers}\sectlabel{quantum}

In spite of difficult conceptual issues~\cite{feynman85,mermin85}, the
properties of quantum mechanics needed to describe the search
algorithm presented in this paper are relatively simple. Specifically,
these properties are the rich set of states available to a quantum
computer and the operations that manipulate these states.

The state of a classical computer can be described by a string of
bits. For example, the complete assignment of an $n$-variable SAT
problem considered at some point in a search is given by $n$
bits. Quantum computers use physical devices whose full quantum state
can be controlled. For example~\cite{divincenzo95}, an atom in its
ground state could represent a bit set to 0, and an excited state for
1. The atom can be switched between these states, e.g., with lasers of
appropriate frequencies. Significantly, the atom can also be placed in
a uniquely quantum mechanical {\em superposition} of these values,
which can be denoted as a vector $(\psi_0, \psi_1)$, with a component
(called an {\em amplitude}) for each of the corresponding classical
states for the system. Similarly, with $n$ such devices, the state of
the quantum machine is described by a vector with $2^n$ amplitudes
$(\psi_0, \psi_1, \ldots, \psi_{2^n-1})$. The amplitudes are complex
numbers that have a physical interpretation: when the computer's state
is measured, the superposition changes, or {\em collapses}, to become
the classical state $s$ with probability $|\psi_s|^2$. Thus amplitudes
satisfy the normalization condition $\sum_s |\psi_s|^2 = 1$. In
contrast to a classical machine which, at any given step of its
program, has a definite value for each bit, the quantum machine exists
in a general superposition of $2^n$ such states.

The change in the superposition brought about by measurement also
means the amplitude values are not explicitly available for use in a
computation. For example, an algorithm that attempts to test the phase
of various amplitudes to determine the next operation would destroy
the superposition. This observation greatly limits the types of
quantum algorithms. Thus using this rich set of states requires
operations that rapidly manipulate amplitudes in a superposition
without explicitly measuring them. Because quantum mechanics is linear
and the normalization condition must always be satisfied, these
operations are limited to unitary linear
operators~\cite{hogg95d}. That is, a state vector $\psi = (\psi_0,
\ldots, \psi_{2^n-1})$ can be changed to a new vector $\psi^\prime$
related to the original one by a unitary transformation, i.e.,
$\psi^\prime = U \psi$ where $U$ is a unitary matrix\footnote{A
complex matrix $U$ is unitary when $U^\dagger U = I$, where
$U^\dagger$ is the transpose of $U$ with all elements changed to their
complex conjugates. Examples include permutations, rotations and
multiplication by phases (complex numbers whose magnitude is one).} of
dimension $2^n \times 2^n$. In particular, this requires that the
operations be reversible: each output is the result of a single
input. In spite of the exponential size of the matrix, in many cases
the operation can be performed in a time that grows only as a
polynomial in $n$ by quantum computers~\cite{boyer96,hoyer97}.

An important example of such operations are reversible classical
programs. Specifically, let $P$ be such a program. Then for each
classical state $s$, i.e., a string of bit values, it produces an
output $s^\prime = P[s]$, and each output is produced by only a single
input. A simple example is a program operating with two bits that
replaces the first value with the exclusive-or of both bits and leaves
the second value unchanged, i.e., $P[00]=00$, $P[01]=11$, $P[10]=10$
and $P[11]=01$.  Denoting the two bits by the binary variables $x$ and
$y$, such a program performs the operation $x \leftarrow x \otimes y$,
where $\otimes$ is the exclusive-or operation. The values of the
variables $x$ and $y$ after this operation uniquely determine the
original values, hence the program is reversible. When used with a
quantum superposition, such classical programs operate independently
and simultaneously on each component, i.e., $P[(\psi_0, \psi_1,
\ldots, \psi_{2^n-1})]$ produces a new superposition $(\psi_0^\prime,
\ldots, \psi_{2^n-1}^\prime)$ where $\psi_{s^\prime}^\prime = \psi_s$
with $s^\prime = P[s]$.  This {\em quantum parallelism} allows a
machine with $n$ bits to operate simultaneously with $2^n$ different
classical states.  Continuing with the 2-bit exclusive-or example, if
we started with the superposition $(1/\sqrt{2},1/2,1/2,0)$ the program
would produce $(1/\sqrt{2},0,1/2,1/2)$. Equivalently, this program
corresponds to the permutation matrix
$$U = \pmatrix{
1 & 0 & 0 & 0 \cr
0 & 0 & 0 & 1 \cr
0 & 0 & 1 & 0 \cr
0 & 1 & 0 & 0 \cr
}$$

These reversible classical programs amount to a permutation of the
states. However, unitary operations can also mix the amplitudes in a
state vector.  An example for $n=1$ is 
\begin{equation}\eqlabel{Umix}
U = \frac{1}{\sqrt{2}}
\pmatrix{ 1 & -1 \cr 1 & 1 \cr }
\end{equation}
This converts $(1,0)$, which could correspond to an atom prepared in
its ground state, to $(1,1)/\sqrt{2}$, i.e., an equal superposition of
the two states.  Because these matrices combine complex numbers, it is
possible for the combination of amplitudes to cancel, leaving no final
amplitude in some of the states. This capability for
interference~\cite{feynman85} distinguishes quantum computers from
probabilistic classical machines.

Designing a quantum search algorithm requires combining these
available quantum operations in a manner that concentrates amplitudes
in solutions.  Classical search techniques can be a useful guide for
such designs. That is, a quantum search method could, at least
approximately, map amplitude from one state to others that would be
considered after it by the corresponding classical method. In effect,
this allows examining, and using interference from, all possible
choices the classical search could have made, rather than being
restricted to a single series of choices. The details of the
particular problem being solved could be introduced by adjustments to
the phases of the amplitudes based on testing states for consistency,
a rapid operation for NP problems.  This technique, used with previous
unstructured~\cite{grover96} and structured~\cite{hogg95d,hogg97}
quantum searches, neatly separates the design of the method that mixes
amplitudes from any consideration of the detailed nature of specific
problems. In the rest of this paper, we show how analogies with local
or repair style searches can be used as the basis for quantum search.

\section{Neighborhood Relations and Quantum Search}

Classical repair style searches consist of randomly selecting a
complete assignment as the initial state, then repeatedly moving to
one of the current state's neighbors until a solution is found or the
number of search steps reaches a prespecified bound. Typically, at
each step the search selects a neighbor that reduces the number of
conflicts, if any such neighbor exists. If the search terminates
without finding a solution, it is repeated from a new random initial
state.

The two key ingredients of these local search algorithms are
identifying the neighboring states (i.e., those with a single
different assignment) and then selecting from among these neighbors
the next one to use. The identification of neighbors doesn't depend on
constraints of the particular problem being solved. On the other hand,
the selection process uses information on the problem instance such as
the number of conflicts in the current state compared to the conflicts
of the neighbors. In some versions of repair searches, such as
simulated annealing~\cite{kirkpatrick83} the selection process also
depends on the number of steps that have been performed (e.g., through
a changing temperature parameter).

The quantum algorithm introduced here operates on superpositions of
all complete assignments for the SAT problem. Furthermore, instead of
selecting only a single new state, each step of the algorithm
considers all possible new states. This parallelism offers the
possibility of improved performance by arranging for interference
among many different classical search paths. A successful algorithm
will adjust amplitudes along these paths so that search paths leading
to solutions tend to combine constructively while those leading to
nonsolutions combine destructively.

We construct quantum analogs of the two key ingredients of classical
repair searches. First, the identification of neighbors corresponds to
an operation on the superposition that moves amplitudes preferentially
from a state to its neighbors. Second, the selection among the
neighbors is performed by adjustments to the amplitude phases, based
on the number of conflicts in the states and their neighbors, i.e.,
the same information that is used in the classical methods.

To describe the resulting algorithm more explicitly, let $\psi^{(
j)}_{s}$ be the amplitude of the assignment $s$ after completing step
$j$ of the algorithm.  A single trial of the search algorithm consists
of the following parts:
\begin{enumerate}
\item initialize amplitude equally among all the assignments, i.e.,
  $\psi^{(0)}_s=1/\sqrt{N}$ for all $s$.

\item iterate: for step $j$ from 1 to $J$, adjust phases and then
  multiply by the matrix $U$ described in \sect{U}, to give
\begin{equation}\eqlabel{map}
\psi^{(j)}_r = \sum_s U_{rs} \rho_s \psi^{(j-1)}_s
\end{equation}
where $\rho_s$ is the phase adjustment for assignment $s$ as described
in \sect{phases}.

\item measure the final superposition
\end{enumerate}

The initialization can be performed rapidly by applying the
matrix of \eq{Umix} separately to each of the $n$ bits.

After $J$ steps, the measurement gives a single assignment,
which is a solution with probability
\begin{equation}\eqlabel{p(soln)}
P_{\rm soln}={\sum_{s}}^\prime p(s)
\end{equation}
with the sum over solutions. Here $p(s)={\left| \psi^{(J)}_{s}\right|
}^{2}$ is the probability to obtain the assignment $s$ with the
measurement. On average, the algorithm will need to be repeated
$T=1/P_{\rm soln}$ times to find a solution.  As with classical repair
searches, this algorithm is incomplete: it can find a solution if one
exists but not prove no solutions exist.  The search cost can be
characterized by the number of steps required to find a solution on
average, i.e., $C = JT$.

Completing the description of the algorithm requires specifying the
number of steps $J$. Because the measurement destroys the
superposition, it is not possible to evaluate $P_{\rm soln}$
explicitly at each step and use this information to determine when to
halt. Instead $J$ must be specified a priori based on readily
available problem characteristics, such as the number of variables $n$
and clauses $m$. The value of $J$ may be changed from one trial to the
next, which allows some exploration for suitable
values~\cite{boyer96}.

In our case, because the algorithm emphasizes mapping amplitude among
neighbors, one might expect that about $n$ steps are required to
ensure an opportunity to move significant amplitude to solutions
(since a solution is at most $n$ neighbors away from any given
assignment).  However, the experiments reported below show fairly
large amplitudes with somewhat fewer steps. In addition, instead of
continuing the algorithm to maximize the probability to have a
solution, a lower average search cost is sometimes possible by
stopping earlier~\cite{boyer96}, a simple strategy for improving
probabilistic algorithms~\cite{luby93}. Determining the number of
steps that minimizes average cost for $k$-SAT problems with given $n$
and $m$ remains an open problem, but at worst one could try the
algorithm for all values of $J$ up to $n$, resulting in a linear
increase in the overall search cost. More sophisticated methods for
finding a suitable number of steps to take have been proposed for the
unstructured search algorithm~\cite{boyer96} and similar techniques
may be useful for this structured search as well.

\subsection{Emphasizing Neighbors with Amplitude Mixing}\sectlabel{U}

For SAT problems with $n$ variables, each state has $n$
neighbors. Thus the matrix $M$ giving the most direct correspondence
with the classical repair searches would have $M_{rs}=0$ whenever the
assignments $r$ and $s$ are not neighbors and $M_{rs}=1/\sqrt{n}$
whenever $r$ and $s$ are neighbors, i.e., $d(r,s)=1$. However, this
matrix is not unitary and so does not correspond to a quantum
operation.

Instead we consider a more general class of matrices whose elements
depend only on the Hamming distance between assignments, i.e.,
$U_{rs}=u_{d(r,s)}$ which can be made unitary by appropriate choices
of the values $u_d$ for $d=0,\ldots,n$. To give a close correspondence
with the classical method, we use the unitary matrix of this form with
the largest possible value of $u_1$, which governs the mapping between
states and their neighbors.

As shown in \app{matrix}, such matrices can be written as $U = W D W$
where, for assignments $r$ and $s$,
\begin{equation}\eqlabel{W}
W_{rs} = \frac{1}{\sqrt{N}} (-1)^\ones{r \bitA s}
\end{equation}
and $D$ is a diagonal matrix of phases (complex numbers with magnitude
equal to 1) depending only on the number of 1-bits in the assignments,
i.e., $D_{rr} = \tau_\ones{r}$. The matrix $W$ is unitary and can be
implemented rapidly on quantum computers using a recursive
decomposition~\cite{boyer96,grover96}. The largest possible value of
$u_1$ is obtained by selecting $\tau_h$ to be 1 for $h \le n/2$ and
$-1$ for $h>n/2$, as shown in \app{matrix}. These values define the
mixing matrix $U$ used in the search algorithm.

For example, when $n=2$ and the states are
considered in binary order, i.e., 00, 01, 10, and 11,
\begin{equation}\eqlabel{U2}
U = \frac{1}{2} \pmatrix{
 1 &  1 &  1 & -1 \cr
 1 &  1 & -1 &  1 \cr
 1 & -1 &  1 &  1 \cr
-1 &  1 &  1 &  1 \cr
}
\end{equation}
For instance, the value mapping the state 00 to its neighbors, 01 and
10, is 1/2, corresponding to the second and third elements of the
first column of this matrix.  

\fig{matrix} shows relative values of the mixing matrix elements for
larger cases with even $n$.  This mixing matrix has a regular pattern
of real-valued entries. Specifically, each column, specifying the
mapping from a particular assignment $s$ to all the others, has a
positive value for assignments $r$ at Hamming distance 0 and 1 from
$s$, negative values for $d(r,s)=2$ and 3, and so on.  More generally,
$u_d$ is negative when $d \bmod 4$ is 2 or 3, and otherwise is
positive.  We make use of this pattern in designing the phase
adjustments, described below.

When $n$ is odd, the pattern for the matrix values is different:
$u_d=0$ for even values of $d$ and the values for odd $d$ alternate in
sign.  That is, $u_d$ is positive when $d \equiv 1 \pmod{4}$, negative
when $d \equiv 3 \pmod{4}$ and otherwise is 0.

\figdef{matrix}{
\epsfig{file=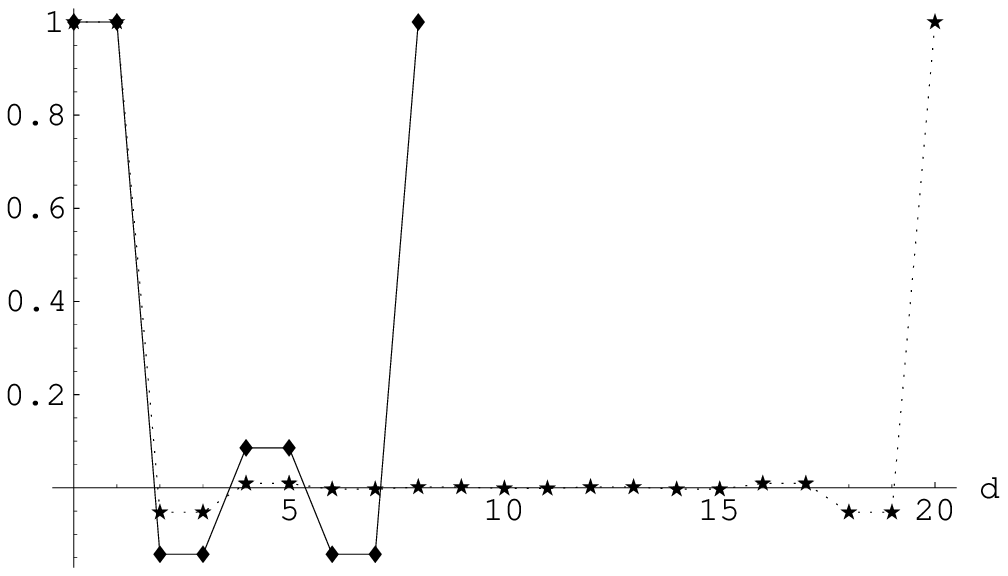}
}{Values of $u_d/u_1$ for $n=8$ (solid) and $n=20$ (dotted). The values
of $u_1$ for $n$ equal to 8 and 20 are 0.27 and 0.18, respectively.}

Unlike the local classical search, there is some mixing between
assignments that are not neighbors.  Generally, these contributions
are relatively small, and become smaller as $n$ increases, for most
assignments.  While these additional connections make it more
difficult to understand the algorithm purely in analogy with classical
repair style searches, the long range couplings provide a possible
mechanism to move amplitude away from local minima, i.e., inconsistent
states all of whose neighbors have more conflicts.

\subsection{Phase Choices}\sectlabel{phases}

The phase adjustment for states, i.e., the values of $\rho_s$ in
\eq{map}, is the portion of the algorithm using information about the
particular SAT problem to be solved. The key to designing an effective
algorithm is selecting phases that, at least on average, move
significant amplitude toward solutions. As with classical heuristics,
the phase selected for a given assignment must be a function only of
readily available information about that assignment and not require
specific knowledge of the problem as a whole, such as the location of
the solutions, that can only be obtained by extensive search.
Furthermore, as described in \sect{quantum}, the phase choices can not
depend on explicit knowledge of the amplitude associated with the
state. For example, the ability to set $\rho_s$ based on the sign of
$\psi_s$ could greatly simplify the design of the algorithm, but such
an operation cannot be physically realized.

Following the analogy with classical local searches, we make use of
the same information such searches use, namely the number of conflicts
in an assignment and how it compares to its neighbors.  We found
several methods that concentrate amplitude into solutions to some
extent.  In this section, we describe two such phase choices. The
first makes few assumptions about the structure of the problem so
applies fairly well to a wide range of problems. The second makes more
specific assumptions about the problem structure and is thereby able
to make more effective use of the pattern of values in $U$, but only
for a limited set of highly constrained problems.

\subsubsection{A Simple Phase Choice}

In the previous structured and unstructured
algorithms~\cite{grover96,hogg95d}, the phase choice consists of
inverting the phase of all inconsistent states, i.e., using $\rho_s=1$
when $s$ is consistent, and otherwise $-1$.  The more recent
structured method~\cite{hogg97} also inverted phases of some
consistent states depending on the step $j$ of the algorithm. In
all these cases, the mapping matrix tended to concentrate amplitude
toward those states whose phases were unchanged. As described in
\app{matrix}, the matrix $U$ of \sect{U} is related to those used these
previous algorithms. Thus we examine using phase inversion as a simple
approach to moving amplitude toward desired states.

In our case, the mapping emphasizes moving amplitude among neighbors.
Since neighbors tend to have similar number of conflicts, we want to
combine the ability of phase inversion to move amplitude toward those
states whose phases are unchanged with an emphasis on neighbors to
take advantage of the large value of $u_1$.  A simple way to do this
is to only invert the phases of states with at least a certain number
of conflicts, and gradually reduce this threshold during the
search. This allows for amplitude to move between neighbors while also
biasing the system toward states with fewer and fewer conflicts.  In
effect this technique attempts to combine amplitude coherently for
states with fewer and fewer conflicts while also giving relatively
large transfers to these states through the large mapping between
neighbor states produced by the matrix $U$. Although this argument
ignores the, sometimes counterproductive, contribution from
non-neighbor assignments, it does serve to motivate the phase
selection method.

Specifically we select an initial threshold $\cStart$ for the
number of conflicts and then at step $j$ the phase for assignment $s$
is
\begin{equation}\eqlabel{phase}
\rho_s = \left\{ \begin{array}{rl}
	-1	& \mbox{if $s$ has more than $\cStart-(j-1)$ conflicts} \\
	1	& \mbox{otherwise}
		\end{array}
	\right.
\end{equation}
The algorithm operates for at most $\cStart+1$ steps (after this
point, \eq{phase} gives the same phase to all states so there are no
further changes to the relative amplitudes).

The exact choice for $\cStart$ is not critical as long as there are
relatively many states with about that number of conflicts. A simple
approach is to start with the average number of conflicts in
assignments, as given by \eq{conflicts}. Thus for our algorithm we
take $\cStart = m/2^k$. Although not used in this paper, a possible
alternative for a given problem, would be to count the number of
conflicts in a set of randomly chosen assignments to estimate the
distribution of conflicts among assignments. This distribution could
in turn suggest a more suitable starting threshold, as well as
different amounts to lower the threshold with each step of the
algorithm.


\subsubsection{Phases Based on Neighborhood Structure}

Another approach to the phase choices is based on the pattern of the
mixing matrix, illustrated in \fig{matrix}. Specifically, if we could
identify a series of disjoint sets of assignments in which successive
sets have a large number of neighbors and the final set consists of
solutions, then we could attempt to move amplitude from one set to the
next based on the large mapping between neighbors. To use this
technique we would also need to be able to initially concentrate
amplitude in the first of these sets.

Identifying such a sequence of sets requires fairly regular
relationships among the number of conflicts of neighboring
assignments. We restrict attention to the case where $n$ is even. A
similar argument can give a phase selection strategy for odd $n$ based
on the different pattern of the $U$ matrix described in \sect{U}.

To illustrate this approach, we consider the simple case of soluble
problems with the maximum possible number of constraints, i.e.,
$m=m_{\rm max}$ from \eq{mMax}. In this case, each assignment with $c$
conflicts has $c$ neighbors with $c-1$ conflicts and the remaining
$n-c$ neighbors have $c+1$ conflicts.  Let $\Nb(s)$ be the number of
neighbors of assignment $s$ that have fewer conflicts than $s$
does. Then for these problems, $\Nb(s) = c$ where $c$ is the number of
conflicts in the assignment $s$.

Let $S_b$ be the set of assignments with $b$ better neighbors, i.e.,
$\{s | \Nb(s)=b\}$. The set $S_0$ contains just the single solution of
the problem. The maximum constrained problems have $n \choose b$
assignments in $S_b$.  Since this quantity is largest when $b=n/2$,
the initial superposition with equal amplitude in all states will have
the largest total contribution for the assignments in $S_{n/2}$. Thus
we will select phases in a way that attempts to move significant
amplitude from the assignments in $S_b$ to those in $S_{b-1}$,
starting with $b=n/2$. If successful, this process will concentrate
amplitude in the solution after $n/2+1$ steps.

We can improve this process by first increasing the amplitude in
the assignments in $S_{n/2}$. This is readily done using the pattern
of signs of the matrix $U$. Specifically, the initial superposition
gives a positive amplitude to every assignment. We can arrange for the
first step of the algorithm to give a sum of positive contributions to
every state in $S_{n/2}$ by selecting the signs of the phases for the
states to equal those of the corresponding matrix elements in $U$. That
is, assignments at Hamming distance $d$ from those in $S_{n/2}$, namely
those in $S_{n/2 \pm d}$, have $\rho_s$ equal to the sign of $u_d$.

For subsequent steps $j>1$, we attempt to move as much amplitude as
possible from assignments in $S_{n/2-(j-2)}$ to those in
$S_{n/2-(j-1)}$. To do this we make use of the fact that the mapping
between neighbors has the largest matrix elements in $U$ and these
elements, $u_1$ are positive. Thus, just considering the contribution
from neighbors, we can expect a large contribution from assignments in
$S_{n/2-(j-2)}$ to those in $S_{n/2-(j-1)}$ by selecting $\rho_s$ for
these assignments to be positive and setting the phase for all other
assignments to $-1$. This argument ignores the contributions from
assignments at larger Hamming distances. However, if a large portion of
the amplitude at step $j$ is concentrated in assignments in
$S_{n/2-(j-2)}$, the other contributions to assignments in $S_{n/2-(j-1)}$
will be small both because the corresponding $u_d$ values are small and
the amplitudes in the other assignments are relatively small.

\begin{mathletters}
\eqlabel{phase1}
This discussion motivates selecting the phase for an assignment $s$ as
follows. For the first step,
\begin{equationGroup}\eqlabel{phase1First}
\rho_s = \left\{ \begin{array}{rl}
	-1	& \mbox{if $\left|\NStart - \Nb(s) \right| \bmod 4 = 2$ or 3} \\
	1	& \mbox{otherwise}
		\end{array}
	\right.
\end{equationGroup}
where $\NStart = \floor{n/2}$.  For subsequent steps, i.e., $j>1$,
\begin{equationGroup}\eqlabel{phase1Rest}
\rho_s = \left\{ \begin{array}{rl}
	1	& \mbox{if $\NStart - \Nb(s) = j-1 $ or $j-2$} \\
	-1	& \mbox{otherwise}
		\end{array}
	\right.
\end{equationGroup}
with the algorithm terminating after at most $j=\NStart+1$ steps.
\end{mathletters}

As suggested by the motivating argument, this choice of phases
requires a close relation between the number of conflicts in an
assignment and its neighborhood relations. While such a relation holds
exactly for the maximum constrained problem, it is only an
approximation for problems with fewer constraints. For example,
problems can have local minima, i.e., assignments for which $\Nb=0$
but which nevertheless have some conflicts and thus are not
solutions. More generally, the neighbors of a given assignment can
have various numbers of conflicts making it more difficult to move
amplitude one step at a time to sets of assignments with successively
fewer conflicts. These characteristics introduce variations in the
amplitudes, in effect causing some randomization of the phases. We
observed the validity of the approximation, and its usefulness as a
search strategy, was lowest for problems around the transition from
over- to underconstrained problems near $m/n=4.2$. Nevertheless, for
highly constrained problems the behavior is close enough to that of
maximum constrained cases to give the step by step shift of a large
portion of the amplitude toward assignments with fewer conflicts.

\section{Quantum Search Behavior}

The behavior of this search algorithm was examined through a classical
simulation. While these results are limited to small problems,
they nevertheless give an indication of how this algorithm can
dramatically focus amplitude into solutions. As a check on the
numerical errors, the norm of the state vector remained within
$10^{-10}$ of 1.

\subsection{Small Example}

We first illustrate the steps in this algorithm, for the trivial
problem with two variables described in \sect{sat}.  For this problem
the characteristics of the assignments and the amplitudes using
\eq{phase} are:
\begin{center}
\begin{tabular}{l|cccc}
  assignment $s$      & 00  &  01  &  10  &  11 \\ \hline
  number of conflicts & 0   &   1  &   1  &   2 \\
  $\psi_s^{(0)}$      & 0.5 & 0.5  &  0.5 & 0.5 \\
  $\rho_s \psi_s^{(0)}$ & 0.5 & 0.5  &  0.5 & $-0.5$ \\
  $\psi_s^{(1)}$      & 1 & 0  &  0 & 0 \\
\end{tabular}
\end{center}
with $c_{\rm start} = 2/2^1 =1$.

Initially the states are prepared with equal amplitudes.  At the first
step \eq{phase} inverts the amplitude of all states with more than one
conflict, i.e., the state 11 in this case. Multiplication by the
mixing matrix of \eq{U2} then gives the final amplitudes. In this
case, after one step, all the amplitude is in the single solution
state so $P_{\rm soln}=1$ and the search cost is $C=1$ step. This
compares with an expected cost of 4 steps from random selection.

The neighborhood based phases choices of \eq{phase1} are based on each
assignment's number of neighbors with fewer conflicts, and $N_{\rm
start}=1$.  For this problem, $|\NStart-\Nb(s)| \leq 1$ so
\eq{phase1First} gives no phase changes for the first step, i.e.,
$\psi_s^{(1)} = \psi_s^{(0)}$. Thus, for this small problem the
attempt to concentrate additional amplitude in assignments with
$\Nb(s) = \NStart$ has no effect. For the next step, $j=2$,
\eq{phase1Rest} inverts the phase for the assignment with $\Nb=2$,
giving
\begin{center}
\begin{tabular}{l|cccc}
  assignment $s$      & 00  &  01  &  10  &  11 \\ \hline
  $\Nb(s)$            & 0   &   1  &   1  &   2 \\
  $\NStart-\Nb(s)$    & 1   &   0  &   0  &   $-1$ \\
  $\psi_s^{(1)}$      & 0.5 & 0.5  &  0.5 & 0.5 \\
  $\rho_s \psi_s^{(1)}$ & 0.5 & 0.5  &  0.5 & $-0.5$ \\
  $\psi_s^{(2)}$      & 1 & 0  &  0 & 0 \\
\end{tabular}
\end{center}
Thus this method also gives $P_{\rm soln}=1$ but only after 2 steps,
so the search cost is $C=2$.

\subsection{Maximum Constrained Problems}

A particularly simple case is given by problems with the maximum
possible number of constraints that still allows for a solution, i.e.,
$m=m_{\rm max}$ as given in \eq{mMax}.

In this case, classical repair style search methods can find a
solution rapidly starting from any initial state, typically by
removing one conflict at a time. Similarly, incremental classical
methods will encounter conflicts immediately upon choosing any
variable value not in the solution, thus allowing the solution to be
found in $O(n)$ steps.

As shown in \app{maxConstraints}, the classical simulation of the
quantum algorithm for these problems runs rapidly.  Thus we are able
to explore the behavior of the algorithm with considerably more
variables than is possible when there are fewer constraints.

The scaling of the search cost for this extreme problem is shown in
\fig{extreme} on a log-log plot where a straight line corresponds to a
polynomial growth in the cost.  The expected search cost grows quite
slowly and is approximately proportional to $n^{1.1}$ over the range
of the figure.  For these problems \eq{phase1} gives somewhat lower
search costs than \eq{phase}. The slow growth in search cost is
particularly impressive since an unstructured quantum search requires
of order $2^{n/2}$ steps, which for $n=100$ is about $10^{15}$. While
this scaling is a polynomial growth in search cost, its cost grows a
bit faster than that of classical heuristics for these problems.

\figdef{extreme}{ 
\epsfig{file=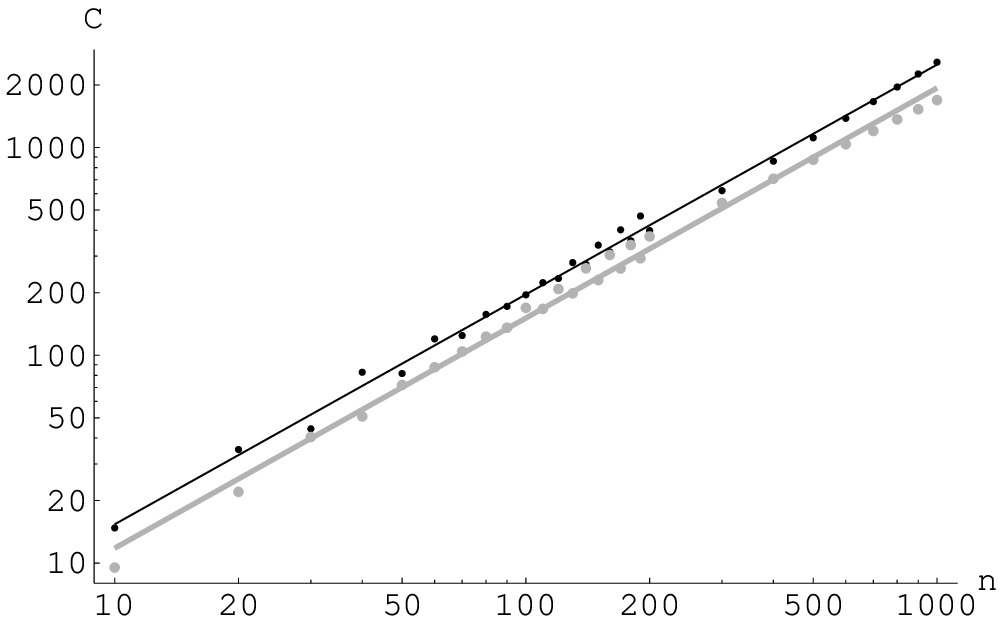} 
}{Expected search cost for maximum constrained 1-SAT vs. $n$ with the
two phase choices: black and gray are \eq{phase} and \eq{phase1},
respectively. The lines are least squares fits to the points.}

For \eq{phase}, the number of steps giving the smallest cost grows
very slowly, ranging from 2 to 4 over the range of variables shown in
the figure. For \eq{phase1}, the largest probability for a solution is
after $n/2+1$ steps. However, for the problems with $n \geq 200$, the
small probability of a solution after just the first step is enough to
give a slightly smaller search cost.

\figdef{amplitude}{ 
\epsfig{file=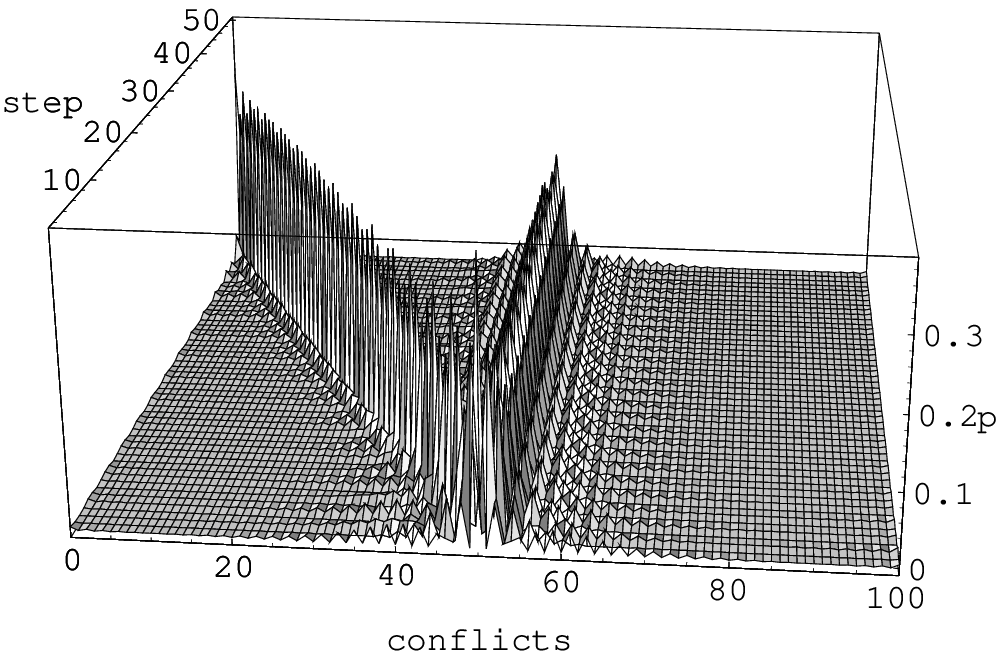}
}{Evolution of the probability in assignments with different numbers of
conflicts for the maximum constrained problem with $n=100$ using
\eq{phase1}.  The probability for a solution reaches about 0.3 after
51 steps, giving a search cost of about 170.
}

Further insight into the behavior of this algorithm is given by
\fig{amplitude} which shows how the probability to have assignments
with different numbers of conflicts varies with each step of
\eq{map}. Specifically, for each step $j$ and each number of conflicts
$c$, the figure shows the value of $\sum_s |\psi^{(j)}_s|^2$ where the
sum is over all assignments $s$ with $c$ conflicts. As described in
\app{maxConstraints}, the amplitudes $\psi^{(j)}_s$ depend only on the
number of conflicts in the assignment $s$. Thus each term in this sum
is the same, giving ${n \choose c} |\psi^{(j)}_c|^2$ where
$\psi^{(j)}_c$ denotes the amplitude of any of the assignments with
$c$ conflicts.  The initial condition (not shown) has equal
probability, $2^{-n}$, in all assignments so the corresponding values
in this plot would be ${n \choose c} 2^{-n}$. We see a large
probability in states with $n/2-(j-1)$ conflicts after step $j$,
illustrating the effectiveness of \eq{phase1} in using neighbor
relations to move amplitude. Moreover, the large spike at 50 conflicts
for step 1 shows the effectiveness of \eq{phase1First} in
concentrating amplitude in states with $\Nb=\NStart$ for this
problem. In this case, ${100 \choose 50} |\psi^{(1)}_{50}|^2 = 0.39$
compared to its initial value ${100 \choose 50} 2^{-100} = 0.08$.

The relatively low search costs and polynomial scaling are also
observed for problems with somewhat fewer constraints than the
maximum, e.g., $m = m_{\rm max}/2$, using this choice of phases.
While this extends the range of good performance for the quantum
algorithm, such a scaling also gives relatively easy problems for
classical heuristics as well.

\subsection{Problems with Fewer Constraints}

The ability of this algorithm to concentrate amplitude into solutions
for highly constrained problems is a significant improvement over
unstructured search methods. However, such problems are inherently
easy because they can be readily solved by classical heuristic
methods, for both incremental and repair style searches.  By contrast,
the difficult search problems, on average, have an intermediate number
of constraints~\cite{hogg95e}: not so few that most assignments are
solutions, nor so many that any incorrect search choices can be
quickly pruned. Specifically, for $k$-SAT the difficult problems have
scaling $m = O(n)$, with the proportionality constant giving the
largest concentration of hard cases depending on the choice of
$k$. For instance, with $k=3$, the hard cases are concentrated
near~\cite{crawford95} $m = 4.2 n$. Thus it is important to examine
the behavior of the algorithm for problems with fewer constraints.
In this section we do so using the phase choice of \eq{phase}.

\figdef{largerExample}{ 
\epsfig{file=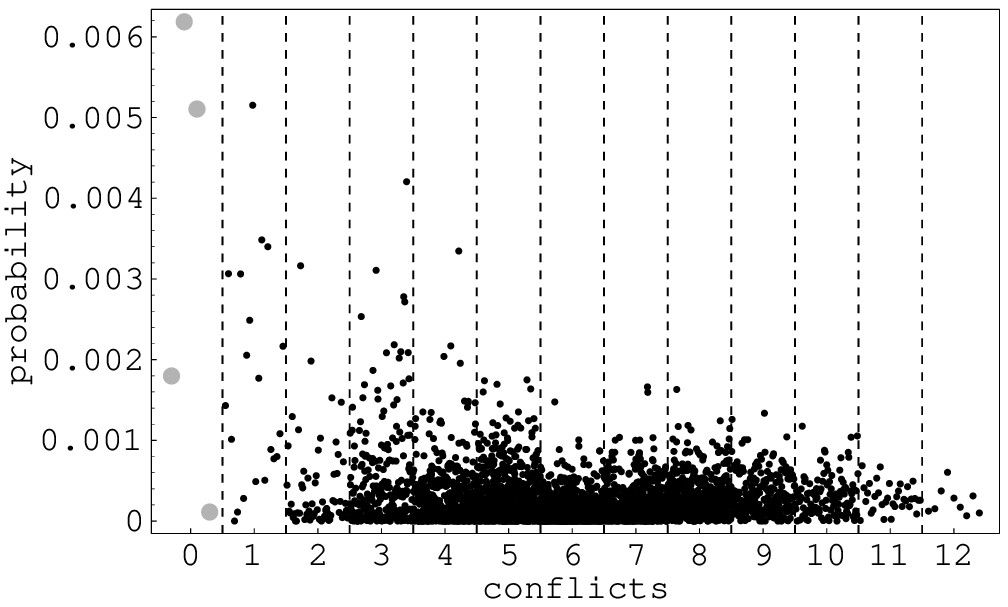} 
}{Probability in each assignment for a 3-SAT instance with $n=12$ after $J=5$
steps. The problem has $m=48$, with 4 solutions and
$P_{\rm soln}=0.013$. The large gray points are the solutions.}

A larger example of how the algorithm concentrates amplitude into
solutions is shown in \fig{largerExample}. It shows the values of
$|\psi^{(J)}_s|^2$ for each of the $2^n$ assignments.  The assignments
are grouped by their number of conflicts and, within each group, by
the integer corresponding to the binary representation of the
assignment.  For clarity, assignments with each number of conflicts
are given the same amount of horizontal space in the plot, even though
there are, e.g., many more assignments with 6 conflicts than with 0 or
12. There are $2^{12}$ assignments, so random selection would give a
probability of about 0.0002 to each assignment. We see that the
algorithm results in many states with relatively few conflicts,
including the solutions, having considerably larger probabilities than
due to random selection.

The figure also illustrates the variation in $|\psi^{(J)}_s|^2$ among
the assignments showing that, unlike the extreme problem of the
previous section, the amplitudes do not depend only on the number of
conflicts. Rather the details of which constraints apply to each
assignment give rise to the variation in values seen here. This
variation precludes a simple theoretical analysis of the algorithm.

\figdef{scaling}{ 
\epsfig{file=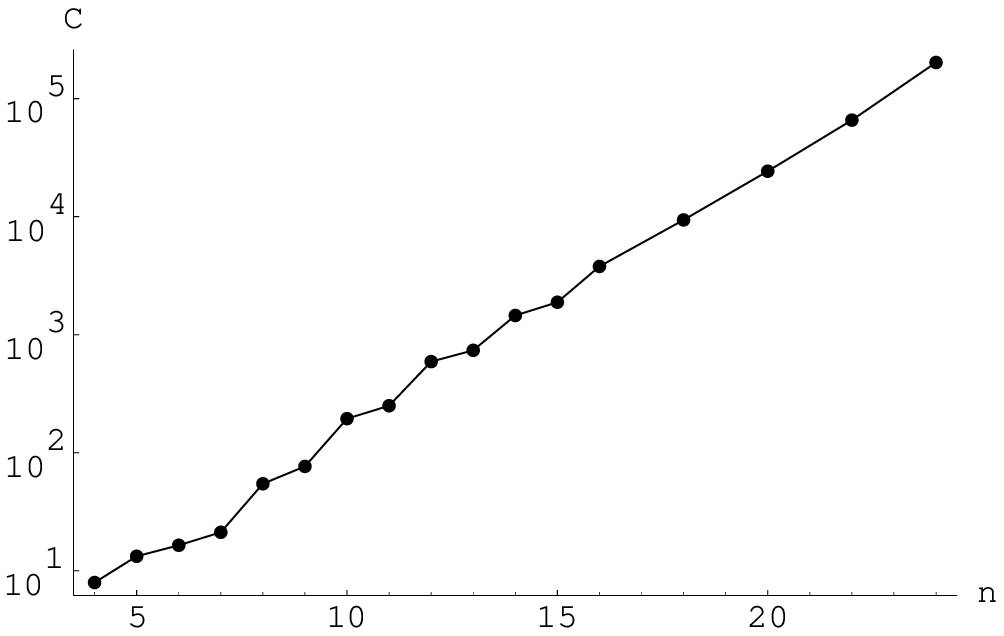} 
}{Expected search cost for random 3-SAT with at least one solution
vs. $n$ with $m/n=4$ on a log plot. For each value of $n$, 1000
problem instances were used. Error bars showing the expected error in
the estimate are included but are smaller than the size of the plotted
points.}

For problems near the transition from over- to underconstrained cases,
the scaling behavior of this algorithm is shown in
\fig{scaling}. Specifically, we generated random soluble instances of
3-SAT problems with $m= 4 n$, as described in \sect{sat}. With $m = 4
n$, a large fraction of the randomly generated instances are soluble,
so random soluble instances are readily generated.  The nearly linear
behavior on this log-plot indicates the seach cost grows exponentially
for these problems. Thus this algorithm is not particularly effective
for the hard problem instances, which are concentrated near the
transition region of $m = 4.2 n$.

\figdef{phase}{ 
\epsfig{file=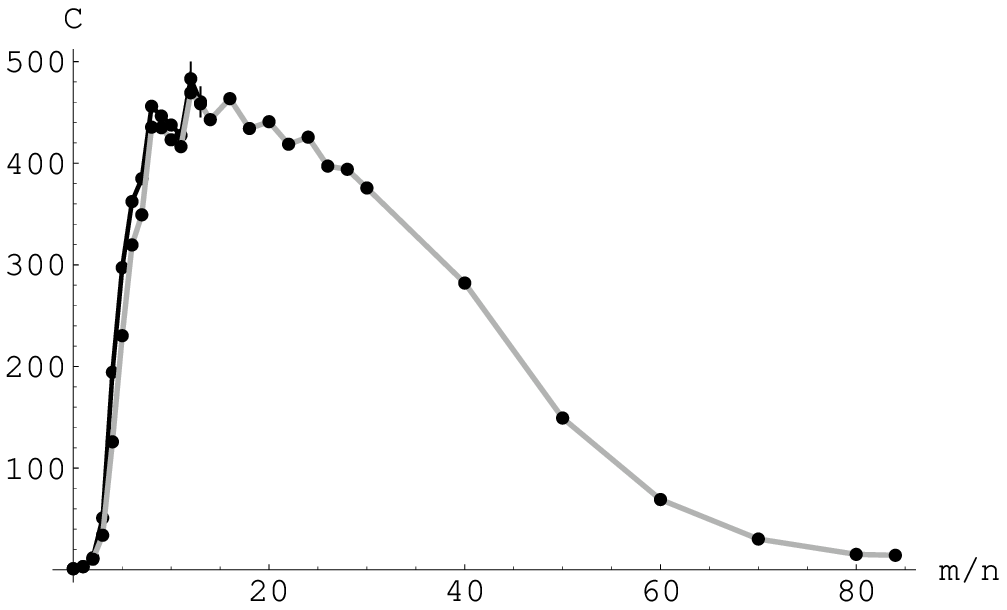} 
}{Expected search cost for random 3-SAT for $n=10$ vs. $m/n$. Each point
represents 1000 problem instances, except only 100 random soluble
instances are shown at $m/n$ equal to 12 and 13. The black line is for
random soluble problems while the gray line is for problems with
prespecified solution. Error bars showing the expected error in the
estimate are included but, in most cases, are smaller than the size of
the plotted points.}

To show how this algorithm is capable of using structure to improve
performance, we examined the behavior as a function of the number of
constraints. To allow sampling the behavior of problems with many
constraints we used random problems with prespecified solution to
cover the full range of $m/n$, generated as described in
\sect{sat}. For comparison, we also examined the behavior of random
soluble problems up to $m/n=15$.  As shown in \fig{phase}, the search
cost eventually decreases as problems become highly constrained,
though at a larger value of $m/n$ than for both the incremental
quantum search method~\cite{hogg97} and classical methods, whose cost
is largest near the transition at $m = 4.2 n$. Thus the local quantum
search method introduced here has only limited effectiveness at using
structure to reduce search cost. Nevertheless, it is better than
unstructured search for highly constrained problems.

\section{ Discussion}

The algorithm presented here shows how the structure of search
problems can be used as the basis of a quantum search algorithm. The
algorithm is particularly effective for relatively highly constrained
problems. It is less effective for problems with an intermediate
number of constraints.

The algorithm might be improved in a number of ways. First, the
initial motivation for the matrix $U$ was to maximally connect
assignments to their neighbors. In fact, we found that the mapping
allowed the algorithm to work best with fewer steps than might be
expected from moving among neighbors one step at a time. It may be
possible to design other mappings that do this even more effectively.

Another issue is the structure of the types of mappings possible with
different choices of the phases for the diagonal matrix $D$. As we
saw, the matrix elements depend only on the Hamming distance between
the assignments when the diagonal elements of $D$ depend only on the
number of 1-bits. It may be helpful to apply the testing operation,
and consequent phase adjustments, more frequently as amplitude is
moved to neighbors than is possible when $U$ has the largest possible
mixing between neighbors. One such method is to decompose $U$ into a
product of matrices, each of which introduces a smaller amount of
mixing, e.g., using $U= W D^{1/s} W$, where $D^{1/s}$ is a diagonal
matrix with $\tau_h=1$ if $h \leq n/2$ and $\tau=e^{i \pi/s}$
otherwise. Phase adjustments can then be introduced between the
components of this product. Another method uses somewhat smaller
values of $u_1$ while retaining real values for $U$ by picking a value
$\alpha <n/2$ and using $\tau_h=1$ if $h \leq \alpha$ and $\tau=-1$
otherwise. The resulting mixing matrix has a more diffusive behavior,
giving smaller changes with each step and hence more opportunity to
apply the phase adjustment. Taken to an extreme, with $\alpha=0$, this
recovers the diffusion matrix of the unstructured
search~\cite{grover96}, which moves too little amplitude among states
at each step to finish rapidly.

There are also a variety of phase adjustment policies. Those studied
here are effective for highly constrained problems, but we found that
other choices can enhance performance in some cases.  Furthermore,
since we focus on typical or average behavior, other choices that do
not improve the average but result in smaller variance would also be
useful in improving the predictablility of the algorithm's
performance.

As a possible extension to this algorithm, it would be interesting to
see whether the nonsolution sets with relatively high probability
could be useful also. If so, the benefits of this algorithm would
be greater than indicated by its direct ability to produce solution
sets. This may also suggest similar algorithms for the related
optimization problems where the task is to find the best solution
according to some metric, not just one consistent with the problem
constraints.

There remain a number of important questions. First, how are the
results degraded by errors and decoherence, the major difficulties for
the construction of quantum
computers~\cite{landauer94a,unruh94,haroche96,monroe96a}? While there
has been recent progress in
implementation~\cite{barenco95,cirac95,cory96,gershenfeld96,sleator95},
quantum approaches to error control~\cite{berthiaume94,shor95} and
studies of decoherence in the context of factoring~\cite{chuang95} it
remains to be seen how these problems affect the algorithm presented
here. However, two aspects of our algorithm may reduce the effect of
errors. First, the algorithm is based on attempts to randomize phases
of contributions to nonsolutions while maintaining similar phases for
contributions to solutions. Since the precise values of the randomized
phases are not critical, some additional variations due to noise
should be tolerable. Second, for problems with $n$ variables the
algorithm requires only $O(n)$ steps during which quantum coherence
must be maintained. That is, coherence need not be maintained between
successive trials of the algorithm. This contrasts with the
$O(2^{n/2})$ coherent steps required by the unstructured
algorithm~\cite{grover96}.  Thus even though our algorithm may need to
be repeated many times to give a good chance of finding a solution, it
has less stringent coherence requirements which can simplify its
hardware implementation.

The second remaining question is due to the algorithm's restriction to
CSPs with two values per variable. This contrasts with the method that
constructs solutions incrementally~\cite{hogg95d} by operating in a
greatly expanded search space. Other CSPs can be converted to
equivalent problems with two values, but it is often better to operate
directly with the problem as specified. Thus an important open
question is whether the neighborhood-based mixing matrix can be
effectively generalized to apply directly to CSPs with larger domain
sizes.

Third, it would be useful to have a theory for asymptotic behavior of
this algorithm for large $n$, even if only approximately in the spirit
of mean-field theories of physics. This would give a better indication
of the scaling behavior than the classical simulations, necessarily
limited to small cases, and may also suggest better phase
choices. Considering these questions may suggest simple modifications
to the quantum map to improve its robustness and scaling.

Finally, there is the general issue of optimally using the information
that can be readily extracted from search states. Local search methods
rely on the number of conflicts and the neighborhood relations among
states. In the method presented here, only a small amount of this
information was actually used to determine the phase
choices. Additional information is available on partial assignments,
as used with incremental searches, but at the cost of involving a
greatly expanded search space.  Making fuller use of this available
information could improve the performance, especially in conjunction
with a theoretical understanding of the typical structure of classes
of search problems~\cite{hogg95e}.

\appendix
\section{Appendix: Neighborhood Mixing Matrix}\sectlabel{matrix}

Here we show that the mixing matrix $U$ used in our algorithm is the
unitary matrix depending only on the Hamming distance between states
that gives the largest possible contribution to neighbors. We do this
in two steps. We first show that matrices whose elements depend only
on Hamming distance can be expressed in the form $W D W$ where $W$ is
given by \eq{W} and $D$ is diagonal with elements depending only on
the number of 1-bits in the assignments. In the second step, we show
that our choice of values for $D$ gives the largest possible mapping
to neighbors among unitary matrices of this form.

\subsection{Matrix Decomposition}

For $n$
variables, the matrices have dimensions $2^n \times 2^n$ and elements
determined by a set of values $u_0,\ldots,u_n$, i.e.,
$U_{rs}=u_{d(r,s)}$. For example, with $n=2$ and the states binary
order, i.e., 00, 01, 10, and 11, the matrix has the general form
\begin{equation}\eqlabel{generalU}
U = \pmatrix{
u_0 & u_1 & u_1 & u_2 \cr
u_1 & u_0 & u_2 & u_1 \cr
u_1 & u_2 & u_0 & u_1 \cr
u_2 & u_1 & u_1 & u_0 \cr
}
\end{equation}

These matrices have a simple recursive decomposition when the states
are in binary order, i.e., ordered by the value of the integer with
corresponding binary representation, namely
\begin{equation}\eqlabel{Udecomposition}
U = 
	\pmatrix{
		U_0 & U_1\cr
		U_1 & U_0\cr
	}
\end{equation}
where the $2^{n-1} \times 2^{n-1}$ matrices $U_0$ and $U_1$ also have
elements depending only on the Hamming distance between assignments,
but considering only the first $n-1$ variables. An example of this
decomposition can be seen in the structure of \eq{generalU}.

This decomposition gives a particularly simple expression for these
matrices. Specifically, define the matrix $\W$ as
\begin{equation}\eqlabel{Wunnormalized}
\W_{rs} = (-1)^\ones{r \bitA s}
\end{equation}
For example, when
$n=2$,
\begin{equation}
\W = \pmatrix{
1 &  1 &  1 &  1 \cr
1 & -1 &  1 & -1 \cr
1 &  1 & -1 & -1 \cr
1 & -1 & -1 &  1 \cr
}
\end{equation}
This matrix also has a
recursive decomposition
\begin{equation}\eqlabel{Wdecomposition}
\W = 
	\pmatrix{
		\W^{\prime} & \W^{\prime}\cr
		\W^{\prime} & -\W^{\prime}\cr
	}
\end{equation}
where $\W^{\prime}$ is the same matrix but defined on assignments to
$n-1$ variables.

Now consider the product $\W U \W$. For $n=1$, this gives
\begin{equation}
\pmatrix{
  2(u_0 + u_1) & 0 \cr
  0            & 2(u_0 - u_1) \cr
}
\end{equation}
which is a diagonal matrix. Now suppose that $\W U \W$ is always
diagonal for any matrix with $U_{rs}$ depending only on $d(r,s)$ up to
$n-1$ variables. Using the recursive decompositions of
\eq{Udecomposition} and \bareeq{Wdecomposition} then gives, for a
matrix with $n$ variables
\begin{equation}
\W U \W = \pmatrix{
  		2\W^{\prime} (U_0+U_1) \W^{\prime} & 0 \cr
		0 & 2\W^{\prime} (U_0-U_1) \W^{\prime}\cr
          }
\end{equation}
Each of the submatrices are diagonal because they involve products of
the form $\W U \W$ for $n-1$ variables. Thus, by induction, $\W U \W$
is a diagonal matrix for all $n$, which we denote by $N D$, where
$N=2^n$. Note that \eq{W} is $W = \W / \sqrt{N}$.  The matrix $W$ is
symmetric and unitary~\cite{grover96}. In particular, $W$ is its own
inverse: $W W = I$.  Thus from $\W U \W = N D$ we obtain $U = W D W$.

Furthermore, the diagonal elements of $D$ depend only on the number of
1-bits in the assignments, i.e., $D_{rr} = \tau_\ones{r}$ for some set
of values $\tau_0,\ldots,\tau_n$. To see this, we have
\begin{eqnarray}
(\W U \W)_{rr}
        & = & \sum_{xy} u_{d(x,y)} (-1)^{\ones{r \bitA x}+\ones{y \bitA r}} \\
        & = & \sum_d u_d \sum_{z=0}^{\ones{r}} (-1)^z {\sum_{xy}}^\prime 1
\end{eqnarray}
where the innermost sum of the second line is over all assignments $x$
and $y$ for which $d(x,y)=d$ and $\ones{r \bitA x}+\ones{y \bitA r} =
z$. For this inner sum, consider those assignments $x$ with exactly
$\eta$ 1-bits in common with $r$, i.e., $\ones{r \bitA x}=\eta$. There
are ${\ones{r} \choose \eta} 2^{n-\ones{r}}$ such choices for
$x$. Then the $y$ sum counts the number of assignments $y$ such that
$d(x,y)=d$ and $\ones{y \bitA r} = z - \eta$. These assignments are
characterized by $a$, the number of 1-bits they share with both $r$
and $x$. The number of such $y$ is readily seen to be
$$
{\eta \choose a} {\ones{r}-\eta \choose z-\eta-a} {n-\ones{r} \choose d-z+2a}
$$
Summing over $a$ then gives the total number of assignments $y$, which
we denote as $\nu(d,z,\eta,\ones{r})$. Combining these observations gives
\begin{equation}
(\W U \W)_{rr} = \sum_d u_d \sum_z (-1)^z \sum_\eta {\ones{r} \choose \eta} 2^{n-\ones{r}} \nu(d,z,\eta,\ones{r})
\end{equation}
which depends only on the number of 1-bits in the assignment $r$.

In summary, {\em any} matrix whose values depend only on the Hamming
distance between assignments can be written in the form $U = W D W$
where $D$ is a diagonal matrix with elements depending only on the
number of 1-bits in the assignments. That is, the values for
$u_0,\ldots,u_n$ are uniquely determined by the diagonal values of
$D$, which in turn are fully specified by the values
$\tau_0,\ldots,\tau_n$ for assignments with each possible number of
1-bits.

Since $W$ is unitary, $U$ will be unitary if and only if $D$ is. For
diagonal matrices, the unitary condition is equivalent to the
requirement that each diagonal element is a phase, i.e., a complex
number whose magnitude equals one.

\subsection{Maximum Mapping to Neighbors}

To determine the choices for $\tau_h$ that give the largest possible
value for $u_1$, we have
\begin{equation}
U_{rs} = \frac{1}{N} \sum_{h=0}^n \tau_h S_h(r,s)
\end{equation}
where
\begin{equation}
S_h(r,s) = \sum_{t, \ones{t}=h} (-1)^{\ones{r \bitA t} + \ones{s \bitA t}}
\end{equation}
with the sum over all assignments $t$ with $h$ 1-bits.

A given 1-bit at position $e$ of $t$ contributes 0, 1 or 2 to $\ones{r
\bitA t} + \ones{s \bitA t}$ when positions $e$ of $r$ and $s$ are
both 0, have exactly a single 1-bit, or are both 1, respectively.
Thus $(-1)^{\ones{r \bitA t} + \ones{s \bitA t}}$ equals $(-1)^z$
where $z$ is the number of 1-bits in $t$ that are in exactly one of
$r$ and $s$. There are $(\ones{r}-\ones{r \bitA s})+(\ones{s}-\ones{r
\bitA s})$ positions from which such bits of $t$ can be selected, and
by \eq{hamming} this is just $d(r,s)$.  Thus the number of assignments
$t$ with $h$ 1-bits and $z$ of these bits in exactly one of $r$ and
$s$ is given by ${d \choose z} {n-d \choose h-z}$ where $d = d(r,s)$. Thus
$S_h(r,s) = S_{hd}^{(n)}$ where
\begin{equation}\eqlabel{Skm}
S_{hd}^{(n)} = \sum_z (-1)^z {d \choose z} {n-d \choose h-z}
\end{equation}
so that $U_{rs}=u_{d(r,s)}$ with $u_d = \sum_h \tau_h S_{hd}^{(n)}/N$.  

To select the values of $\tau_h$ that maximize $u_1$, note that
$S_{01}^{(n)} = 1$ and $S_{h1}^{(n)} = {n-1 \choose h} - {n-1 \choose
h-1}$.  Thus $S_{h1}^{(n)}$ is positive for $n>2h$ and negative for
$n<2h$, and $u_1$ is maximized by selecting $\tau_h$ to be 1 for
$h<n/2$ and $-1$ for $h>n/2$. If $n$ is even, $S_{h1}^{(n)}$ is zero
for $h=n/2$ so the choice of $\tau_{n/2}$ does not affect the value of
$u_1$.  In this case, we take $\tau_{n/2}=1$.  These choices give $u_1
= \frac{2}{N} {n-1 \choose \lfloor n/2 \rfloor}$ which scales as
$\sqrt{2/(\pi n)}$ as $n \rightarrow \infty$. Note this is much larger
than the off-diagonal matrix elements in the diffusion matrix used in
the unstructured search algorithm~\cite{grover96}, which are $O(1/N) =
O(2^{-n})$.

This maximum neighbor mixing matrix is identical to that used with an
incremental search method~\cite{hogg97} but with a different
interpretation of the search states. If instead we take $\tau_0=-1$ and
the remaining values equal to one $U$ becomes the diffusion matrix of
the unstructured search method~\cite{grover96}.

\section{Appendix: Classical Simulation}

A direct classical simulation of the quantum algorithm results in an
exponential increase in the cost, limiting the empirical evaluation of the
algorithm to small cases. Nevertheless, the simple structure of the mixing
matrix $U$ can be used to reduce this cost penalty. It can be further
reduced for the special case of soluble problems with the maximum
possible number of constraints.

\subsection{General Problems}\sectlabel{classical}

As a practical matter, it is helpful if a quantum search method can be
evaluated effectively on existing classical computers.  Unfortunately,
the exponential slowdown and growth in memory required for such a
simulation severely limits the size of feasible problems. For example,
\eq{map} is a matrix multiplication of a vector of size $2^n$ so a
direct evaluation requires $O(2^{2n})$ multiplications.

The cost of the classical simulation can be reduced substantially by
exploiting the map{'}s simple structure.  Specifically, the product
$\W{\bf x}$ can be computed recursively using \eq{Wdecomposition}
giving
\begin{equation}
\W{\bf x} =
	\pmatrix{
		\W^{\prime}{\bf x^{(1)}} + \W^{\prime}{\bf x^{(2)}} \cr
		\W^{\prime}{\bf x^{(1)}} - \W^{\prime}{\bf x^{(2)}} \cr
	}
\end{equation}
where $\bf x^{(1)}$ and $\bf x^{(2)}$ denote, respectively, the first
and second halves of the vector $\bf x$. Thus the cost to compute
$\W{\bf x}$ is $C(n)=2C(n-1)+O(2^n)$ resulting in an overall cost of
order $n2^n$ for this product as well as the full mapping step $U{\bf
x} = (1/N) \W D \W{\bf x}$. While still exponential, this improves
substantially on the cost for the direct evaluation on classical
machines. An open question is whether other techniques could give
approximate values for the behavior without an exponential cost on
classical machines~\cite{cerf97}.

\subsection{Maximally Constrained Problems}\sectlabel{maxConstraints}

In general, the cost of a classical simulation of the quantum
algorithm grows exponentially with the number of variables. However,
the simulation cost is greatly reduced in some special cases where the
amplitudes have a regular structure. One such case is provided by
problems with the maximum possible number of constraints that still
allows for a solution.

For such problems, the amplitudes depend only on the number of
conflicts in each assignment, not their particular location. This
observation allows for a smaller representation of the search state
and hence an empirical study of larger cases.  To see this, consider a
1-SAT problem with $n$ variables. The most constrained, but still
soluble, case will have $n$ conflicts, each involving a distinct
variable. Thus each variable has an associated ``bad bit'' value that
causes a conflict. The number of conflicts in a given assignment $s$
is then just equal to the number of bad bits it contains. An
assignment with $c$ conflicts will have $c$ neighbors with $c-1$
conflicts and $n-c$ neighbors with $c+1$ conflicts. Thus the
neighborhood structure of the problem is uniquely determined by the
number of conflicts in the assignments.

Suppose $\psi^{(j)}_s$ depends only on the number of conflicts in the
assignment $s$. We need to show that after a single step, \eq{map}
gives values for $\psi^{(j+1)}_r$ that only depend on the number of
conflicts in $r$. First note that the phase choice $\rho_s$ uses only
the neighborhood structure of $s$, which for these problems depends
only on the number of conflicts in $s$. Thus $\rho_s \psi^{(j)}_s$
depends only on $c$, the number of conflicts in $s$, and can be
denoted by $\phi_c$. Since the mixing matrix $U_{rs}$
depends only on the Hamming distance between the assignments $r$ and
$s$, \eq{map} becomes
\begin{equation}
\psi^{(j+1)}_r = \sum_d u_d \sum_c \phi_c {\sum_s}^\prime 1
\end{equation}
where the inner sum is over assignments $s$ with $c$ conflicts and
Hamming distance $d$ from the assignment $r$. Suppose $r$ has $b$
conflicts. Of the $c$ ``bad bits'' in $s$, $(c+b-d)/2$ are in common
with those of $r$, and the remaining $(c-b+d)/2$ do not appear in
$r$. The number of ways to construct such assignments $s$ is then
$$ {b \choose (c+b-d)/2} {n-b \choose (c-b+d)/2} $$ Thus the result of
this mapping step depends only on the number of conflicts in the
assignment $r$, preserving the simple structure of the search state.
Thus we can represent the entire search state by simply keeping track
of the amplitudes associated with each possible number of conflicts,
i.e.,
\begin{equation}
\psi^{(j+1)}_b = \sum_c V^{\rm max}_{bc} \rho_c \psi^{(j)}_c
\end{equation}
where $b$ and $c$ range from 0 to $n$ and
\begin{equation}
V^{\rm max}_{bc} = \sum_d u_d {b \choose (c+b-d)/2} {n-b \choose (c-b+d)/2}
\end{equation}

Because the search state depends only on the number of conflicts and
not their specific associated variables or values, we can, without
loss of generality, choose the conflicts so that all the ``bad bits''
have the value 1, i.e., the unique solution of the problem has all
variables equal to 0. With this choice, we can then make use of the
decomposition of the mixing matrix. That is, $V^{\rm max} = W^{\rm
max} D^{\rm max} W^{\rm max}$ where $D^{\rm max}$ is a diagonal matrix
with $D^{\rm max}_{bb}$ equal to 1 for $b \leq n/2$ and $-1$
otherwise, and
\begin{equation}
W^{\rm max}_{bc} = \frac{1}{\sqrt{N}}\sum_z (-1)^z {b \choose z} {n-b \choose c-z} = \frac{S_{cb}^{(n)}}{\sqrt{N}}
\end{equation}
from \eq{Skm}.  Here the binomials in the sum count, for an assignment
$r$ with $b$ 1-bits, the number of assignments $s$ with $c$ 1-bits
that have $\ones{r \bitA s}=z$. This decomposition is possible for
this particular choice of the unique solution because the number of
1-bits correspond to the number of conflicts.

In this way, classical evaluation of \eq{map} reduces to
multiplication by matrices of size $(n+1) \times (n+1)$, with cost of
order $n^2$.

Finally, 1-SAT problems with fewer than the maximum number of
constraints also have the property that the amplitudes depend only on
the number of conflicts in each assignment. So this compact
representation could be used to study other cases of 1-SAT with many
more variables than is feasible for a direct classical simulation.

\end{document}